\documentclass[prb, twocolumn,showpacs,preprintnumbers,amsmath,amssymb]{revtex4}

\usepackage{graphicx}
\usepackage{dcolumn}
\usepackage{multirow}
\usepackage{bm}
\usepackage{epstopdf}

\newcommand{\beq}{\begin{equation}}
\newcommand{\eeq}{\end{equation}}
\newcommand{\beqs}{\begin{eqnarray}}
\newcommand{\eeqs}{\end{eqnarray}}

\def\hbar{\hspace{0pt}\raisebox{1pt}{$-$} \hspace{-7pt} h}

\begin{document}

\title{Efficient method to calculate total energies of large nanoclusters}

\author{M. Yu$^{1,3}$, R. Ramprasad$^{2}$,  G. W. Fernando$^{1}$ and Richard M. Martin$^{3}$}
\affiliation{$^1$Department of Physics, University of Connecticut,
Storrs, CT 06269}
\affiliation{$^2$Department of Chemical,
Materials and Biomolecular Engineering, University of Connecticut,
Storrs, CT 06269}
\affiliation {$^3$Department of Physics,
  University of Illinois at Urbana-Champaign, IL 61801}

\date{\today}

\begin{abstract}
We present an approach to calculate total energies of nanoclusters based on first principles estimates. For very large clusters the total energy can be separated into surface, edge and corner energies, in addition to bulk contributions. Using this separation and estimating these with direct, first principles calculations, together with the relevant chemical potentials, we have calculated the total energies of Cu and CdSe tetrahedrons containing a large number of atoms. In our work we consider polyhedral clusters so that in addition our work provides direct information on relaxation. For Cu the effects are very small and the clusters vary uniformly from very small to very large sizes. For CdSe there are important variations in surface and edge structures for specific sizes; nevertheless, the approach can be used to extrapolate to large non-stoichiometric clusters with polar surfaces.
\end{abstract}


\maketitle

\section{Introduction}

Nanoscience provides an ideal platform in the search for novel
materials with desirable and tunable properties.
Nanocrystals (NCs) of various sizes and shapes have been found to
exhibit a wide variety of physical and chemical
 properties~\cite{Murray, Alivisatos, Skolnick & Mowbray, Law}, rarely
seen in bulk materials. Synthesizing nanoparticles of a given size
and shape is notoriously difficult and has become a key focus area
due to technically significant properties that depend on the
size/shape of the cluster. Growth of such clusters is governed by
both kinetics and thermodynamics~\cite{Tiller}. If a NC has a
highly symmetric crystal structure (such as zincblende), it is
likely that when synthesizing, the crystal will grow with no
preferred direction of growth. On the other hand, if the crystal
structure has a preferred axis of symmetry, as in the hexagonal
wurtzite structure which has a unique polar axis, preferential
growth along this axis can be expected~\cite{Peng, Nirmal}. Energies associated with
various facets (i.e., with different surface
orientations) will also play a key role during growth. The total
energy, which depends on these facet orientations,
will determine the stability of a nanocluster having a
given size and shape~\cite{Cleveland & Landman, Yacaman, Zhang}. Calculating these energies associated with
large nanocrystals is a nontrivial task. Traditional ``brute force"
or direct first principles methods become quite laborious as the
size of the cluster increases.

First principles methods have provided reliable total energies for
atoms and molecules; this is also true in solids, provided that
the number of atoms in a given unit cell is relatively small. When this
number becomes large (say more than a thousand), then it becomes
computationally prohibitive to carry out first principles
calculations. Naturally, it is desirable to
develop alternate techniques to obtain quality total energies
of systems containing a large number of atoms. This
work is focused on obtaining such total energy estimates using
first principles values calculated for a small number of
well defined clusters.  This approach provides crucial information
needed to treat accurately these clusters as well as general results
that can be used for other classes of clusters.

\section{Methodology}

\subsection{Calculational method for clusters}

Surface energies have been extracted from first principles slab
(total energy) calculations by several groups (see for example, Refs.~\cite{Manna,Bernholc}).
Besides that, a first principles wedge-shaped approach has also
been proposed to calculate the surface energies of polar surfaces~\cite{Zhang}.
In the present work, our goal is somewhat different compared to the above.
We will use energies obtained from first principles calculations
on polyhedral nanoclusters to estimate total energies of larger nanoclusters.
The energy can be written as
\begin{eqnarray}
  E_{tot}^{poly} = \sum_{\alpha}\sigma_{\alpha}A_{\alpha} +
 \sum_{\beta}\epsilon_{\beta}L_{\beta}
     + \sum_{\gamma}c_{\gamma} + \sum_{i}N_{i} \mu_{i},
\label{eq:etot}
\end{eqnarray}
where, $\sigma_\alpha$, $\epsilon_\beta$ and $c_\gamma$ denote
surface, edge and corner energies respectively, $N_{i}$ is the
number of atoms of type $i$,
and $\mu_i$ is the corresponding chemical potential.
The last term in Eq.~(\ref{eq:etot}) contains the bulk energy
if the chemical potentials satisfy the conditions given in section II B.
For a stoichiometric cluster the total energy is independent
of the individual chemical potentials
and is well-defined relative to the energy of the bulk crystal.

In the present work,
a least-squares fit for the total energies of several small clusters
were used to estimate the surface, edge and corner energies (as parameters).
If the above energies can be evaluated in a computationally
efficient way, then the total energy of a cluster (polyhedron),
containing a (substantially) large number of atoms, can be expressed
 {\sl algebraically} as in Eq.~(\ref{eq:etot}).
We demonstrate that using accurate density
functional theory (DFT) based estimates of total energies of a few
small clusters, it is possible to estimate the above (surface,
edge, corner) contributions (as well as chemical potentials) and
then use them (as parameters) in larger clusters to estimate
total energies. The calculations were based on the local density
approximation within the DFT~\cite{Martin} as implemented in the
local orbital SIESTA code~\cite{Siesta}. Norm-conserving nonlocal
pseudopotentials of the Troullier-Martins type~\cite{Troullier & Martins}
were used to describe all the elements.

For a specific shape, the cluster can be defined by one characteristic length $\ell L_{0}$,
so that the surface and edge terms can be expressed  as
$A_{\alpha}=a_{\alpha}\ell^{2}$,
$L_{\beta}=b_{\beta}\ell^{1}$ respectively,
with $a_{\alpha}$ and $b_{\beta}$ being constants.
Therefore, Eq.~(\ref{eq:etot}) may be written in the following form;
\beqs
&&(E_{tot}^{poly} - \sum_{i}N_{i} \mu_{i}) / \ell^{2} \nonumber \\
&& = \sum_{\alpha}\sigma_{\alpha}a_{\alpha} + \sum_{\beta}\epsilon_{\beta}b_{\beta}/\ell
     + \sum_{\gamma}c_{\gamma}/\ell^{2}.
\label{eq:etot_l2}
\eeqs
We label the energy expression on the left side of Eq.~(\ref{eq:etot_l2}) as
``termination energy'' since it is the added
energy due to the presence of surfaces, edges and corners.
For example, for a tetrahedral structure bounded by four $(111)$ surfaces.
(see Fig.~\ref{fig:cdse_zb}), The characteristic size is $\ell L_0$
($\ell$ being an integer and $L_0$  the nearest neighbor distance along an edge),
representing an edge of a triangular (111) facet.
The tetrahedral structure includes four equivalent
$(111)$ surfaces, six equivalent $(111)-(111)$ edges, and four
equivalent $(111)-(111)-(111)$ corners. This choice enables us to
work with well defined (111) surfaces, as well as equivalent
corners, edges and surfaces. In this case, the termination energy in
Eq.~(\ref{eq:etot_l2}) turns out to be
\begin{eqnarray}
(E_{tot}^{poly} -\sum_{i}N_{i}\mu_{i})/\ell^{2}  &=&
\sqrt{3}\sigma L_{0}^{2} + 6\epsilon L_{0}\frac{1}{\ell} + 4c\frac{1}{\ell^{2}}.
\label{eq:etot_tetra}
\end{eqnarray}
In later discussions, we use the above equation for both FCC Cu clusters
and zincblende CdSe clusters.

\subsection{Non-stoichiometric case}

One of the examples we have chosen to study is the binary compound CdSe.
When the cluster is stoichiometric, the number of Cd atoms will be equal
to the number of Se atoms, and the sum of the chemical potentials,
$\mu_{Cd} + \mu_{Se}$ = total energy per CdSe pair,
can easily be evaluated from bulk total energy calculations.
In a non-stoichiometric case, the number of Cd atoms is different from
the number of Se atoms, and some energy terms, such as the surface
energies, depend on the separate values of $\mu_{Cd}$ or $\mu_{Se}$,
instead of their sum. To examine this further,
we set lower and upper bounds for the individual
chemical potentials (pertaining to  bulk CdSe) as
\beqs
 E_{tot}^{bulk}(CdSe) &=& \mu_{Cd} + \mu_{Se} \label{eq:chempot}\\
        &=& E_{tot}^{bulk}(Cd) + E_{tot}^{bulk}(Se) + \Delta H_{f}, \label{eq:formation}
\eeqs
where $\Delta H_{f}$ is the formation energy.
The bulk values, $E_{tot}^{bulk}(Cd)$ and $E_{tot}^{bulk}(Se)$,
are obtained from total energy calculations of pure Cd and pure Se
(per atom, in their equilibrium structures) separately,
while $E_{tot}^{bulk}(CdSe)$ is the total energy of a CdSe pair in the bulk.

Following well known procedures~\cite{Martin}, we can set  bounds listed
below for the chemical potentials of the individual species in CdSe:
\beqs
E_{tot}^{bulk}(Cd) + \Delta H_{f} \leq \mu_{Cd} \leq E_{tot}^{bulk}(Cd)\label{eq:bounds1}\\
E_{tot}^{bulk}(Se) + \Delta H_{f} \leq \mu_{Se} \leq E_{tot}^{bulk}(Se).\label{eq:bounds2}
\eeqs
The right hand side of the first inequality represents the fact that,
$\mu_{Cd}$ in the cluster must be smaller than the (pure Cd)
bulk value $E_{tot}^{bulk}(Cd)$, since otherwise, Cd must phase separate.
The left hand side of this inequality follows from the fact that for $\mu_{Se}$,
one can use the same argument and utilize Eq.~(\ref{eq:formation}) to obtain the following:
\beqs
 0 \leq  - \mu_{Se} + E_{tot}^{bulk}(Se)
      = \mu_{Cd} - E_{tot}^{bulk}(Cd) - \Delta H_{f}.
\label{eq:form2}
\eeqs

From the above arguments, it appears that we can only evaluate the individual
chemical potentials, $\mu_{Cd}$ and $\mu_{Se}$, within the
range given above. However, for a specific family of  non-stoichiometric
clusters having the same shape, we demonstrate below that
the total energy can be determined with no knowledge of the chemical potentials.
\begin{figure}[htp]
 \begin{center}
  \includegraphics[width=15pc]{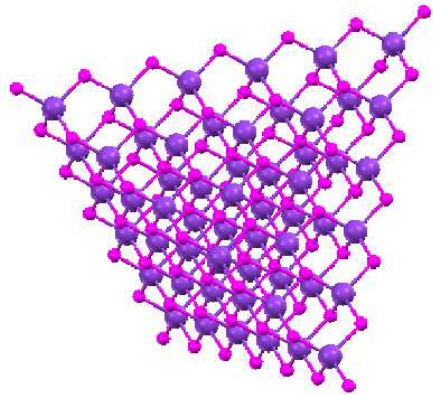}
   \caption{\small CdSe Zincblende (tetrahedral) structure bounded by four $(111)$ surfaces
terminated by Cd atoms. Small dots represent Cd atoms,
while large dots represent Se atoms. The characteristic size is $\ell L_0$
($\ell$ being an integer, with $\ell=6$  shown here), representing an edge of a triangular
(111) facet. } \label{fig:cdse_zb}
 \end{center}
\end{figure}

We first show that the total energy of any CdSe (zincblende)
tetrahedron, $E_{tot}^{poly}$, bounded by similar (111) facets as shown in
Fig.~\ref{fig:cdse_zb}, can be expressed in a slightly different
form of Eq.~(\ref{eq:etot}), i.e.,
\beqs
&&E_{tot}^{poly}(CdSe)-N_{Cd}\mu_{Cd}-N_{Se}\mu_{Se} \nonumber \\
&& = \sqrt{3}\sigma L_{0}^{2}\ell^{2} + 6\epsilon L_{0}\ell + 4c.
\label{eq:cdse0}
\eeqs
Here $L_0$ is the nearest neighbor distance along an edge of the tetrahedron.

Combining with Eq.~(\ref{eq:chempot}), we obtain:
\beqs
&&E_{tot}^{poly}(CdSe)-N_{Se}E_{tot}^{bulk}(CdSe) \nonumber \\
&& = (N_{Cd}-N_{Se})\mu_{Cd} +
\sqrt{3}\sigma L_{0}^{2}\ell^{2} + 6\epsilon L_{0}\ell + 4c.
\label{eq:cdse} \eeqs
However, note that
\beqs
N_{Cd}-N_{Se} &=& \frac{1}{2}\ell^{2}+\frac{3}{2}\ell+1,
\eeqs
which results from a simple count of the atoms in a tetrahedron
having an edge of length $\ell L_0$ ($\ell$ being an integer
and $L_0$  the nearest neighbor distance along an edge).
This result clearly shows that the difference in the number of Cd and Se atoms
arises from surfaces ($\ell^2$), edges ($\ell^1$) or corners ($\ell^0$),
and leads to the following important simplification:
\beqs
&&E_{tot}^{poly}(CdSe)-N_{Se}E_{tot}^{bulk}(CdSe) \nonumber \\
&& = (\frac{1}{2}\mu_{Cd}+\sqrt{3}\sigma L_{0}^{2})\ell^{2} +
(\frac{3}{2}\mu_{Cd}+6\epsilon L_{0})\ell+(\mu_{Cd}+4c). \nonumber \\
\label{eq:CancelVolTerm}
\eeqs

The significance of Eq.~(\ref{eq:CancelVolTerm}) is that,
even in this non-stoichiometric case, it is possible to estimate the total energies
independent of the chemical potentials. This is because in the above equation,
the coefficients of $\ell^2$, $\ell$ and the constant term
act as straightforward parameters to be estimated.  Since there is
no volume ($L^3$) term on the right side of Eq.~(\ref{eq:CancelVolTerm}),
neither $\mu_{Cd}$ nor $\mu_{Se}$ will have a direct effect on the final
total energy to be predicted.
Note that $E_{tot}^{bulk}(CdSe)$ has
an unambiguous value, as the total energy per $CdSe$ pair in the (bulk) zincblende
structure.
Now the parameters, such as surface energy, edge energy, and corner energy,
can be fitted using several, known (DFT based) total energy values from small clusters.
Finally, in order to predict the total energies of  large polyhedrons,
the algebraic expression in Eq.~(\ref{eq:CancelVolTerm}) can be utilized as previously.

\section{Results }

Test results from pure (fcc) Cu as well as (zincblende) CdSe
clusters show that this scheme is reliable to a high degree of
accuracy. One of the significant results of the present study is
our ability to calculate energies of nanoclusters that are
non-stoichiometric and that have polar surfaces. Furthermore, we
will demonstrate that the total energies can be evaluated
independently of individual chemical potentials
(Eq.~(\ref{eq:bounds1})~(\ref{eq:bounds2})).

\subsection{ FCC Cu}

We begin our discussion with pure fcc Cu clusters, which are
regular tetrahedral clusters having a characteristic length,
$L=\ell L_{0}$ ($\ell$ being a positive integer, $L_{0}=2.545\AA$
after geometric relaxation), bounded by four, equivalent (111) facets.
The relevant chemical potential,
$\mu_{Cu}$, can be obtained from a fcc bulk total energy
calculation. The first principles total energies, $E_{tot}^{poly}$, for
the tetrahedrons are obtained after fully relaxing a given cluster.

\begin{figure}
 \begin{center}
  \includegraphics[width=21pc]{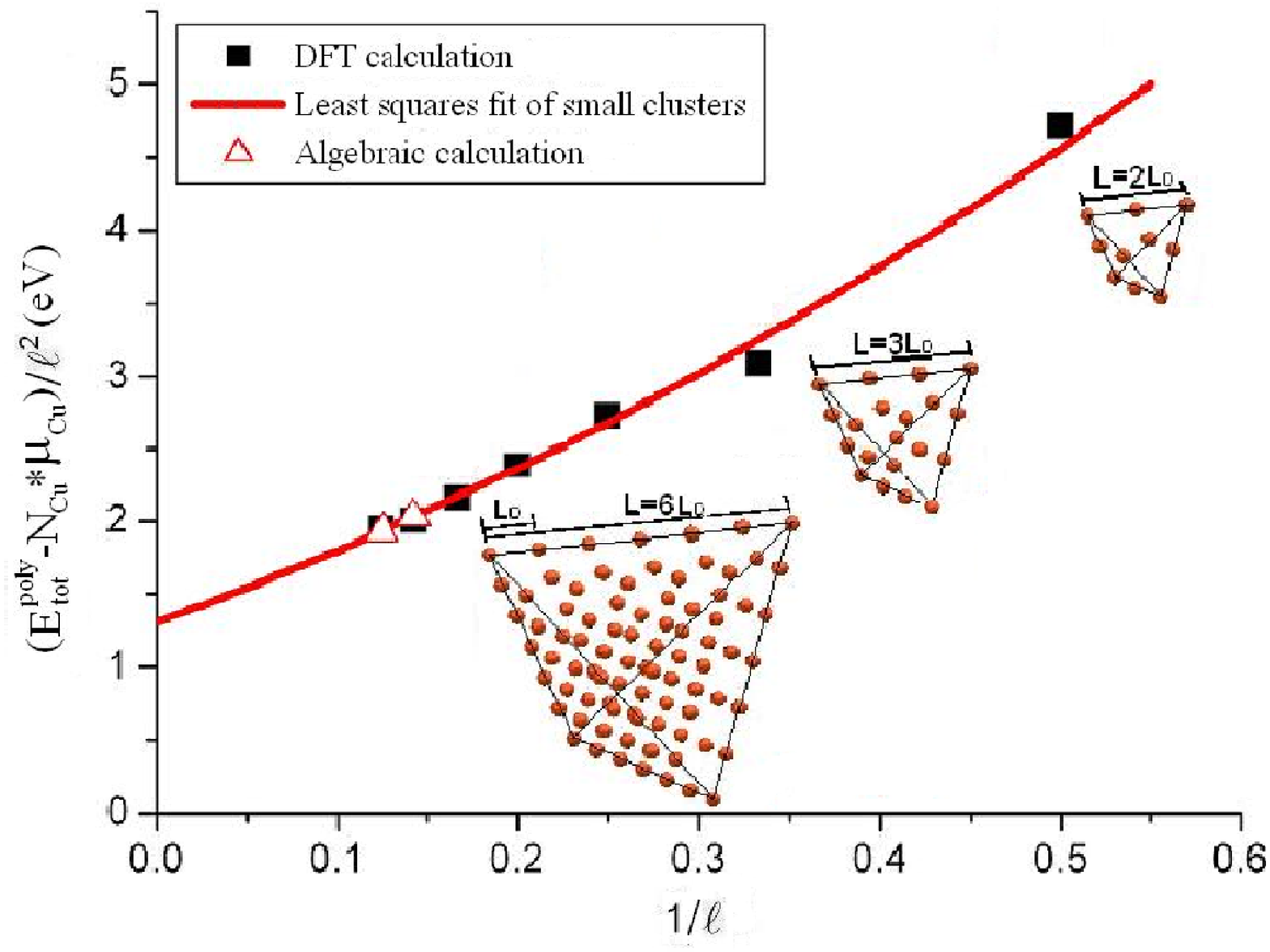}
   \caption{\small The termination energy (see text for definition)
 contribution scaled by $\ell^{2}$ for Cu clusters as a function of the inverse
 of the characteristic size ($1/\ell$) of the tetrahedrons.
 Y-intercept is proportional to surface energy($\sigma$), slope is proportional
 to edge energy($\epsilon$), and curvature is proportional to corner energy($c$).
 The solid squares represent the values calculated using DFT while
 the open triangles represent the estimated values from our algebraic expression.
 The solid line is a least squares fit for the DFT points. }
\label{fig:cu_etot}
 \end{center}
\end{figure}

We find  it is more useful to focus on the energy term
$(E_{tot}^{poly} - N_{Cu}\mu_{Cu} )$, since this represents a termination energy
to the bulk chemical potential contribution
due to the presence of surfaces, edges and corners.
The termination energies scaled by the square of a
characteristic length $\ell^2$, i.e., $(E_{tot}^{poly} - N_{Cu}\mu_{Cu} )/\ell^{2}$,
for different sized clusters ($\ell=2$ to $\ell=8$) calculated using DFT are shown
as solid squares in Fig.~\ref{fig:cu_etot}.
To estimate surface, edge and corner energies, total energies of a
few small clusters ($\ell=2,3,4,5,6$) were used in a least squares
fit according to Eq.~(\ref{eq:etot_tetra}). The estimated $\sigma$,
$\epsilon$ and $c$ were $0.1164~eV/\AA^{2}$, $0.2949~eV/\AA$, and
$1.0047~eV$ respectively. A previous DFT calculation has reported
the (111) fcc Cu surface energy
$\sigma = 0.1213 eV/\AA^{2}$ ~\cite{Scheffler} while the
experimental result for the surface energy of $Cu_{(111)}$ is $0.1144
eV/\AA^{2}$ ~\cite{Boer}.
Hence our calculated $\sigma$ value is in good agreement with previous
theoretical and experimental values.
Using the three parameters for surface, edge and corner energies, we  now
estimate the termination energy as a function of cluster size,
as shown in Fig. \ref{fig:cu_etot}.
The estimated total energies for the two larger
clusters ($\ell=7$ and $\ell=8$) are carried out using the
algebraic expression shown in Eq.~(\ref{eq:etot_tetra}).

\begin{table}
{\begin{center}
\begin{tabular}{||c|c|c|c||}
\hline
\hline
\multirow{2}{*}{$\ell~(L_0)$}&\multirow{2}{*}{$N_{Cu}$}&\multicolumn{2}{c||}{$E^{poly}_{tot}/\ell^{2}~(eV)$} \\
\cline{3-4}
                                                                        & &DFT&algebraic \\
\hline
 2 & 10 & -3041.43 &  \\
 3 & 20 & -2704.60 & \\
 4 & 35 & -2662.65 & \\
 5 & 56 & -2726.96 & \\
 6 & 84 & -2840.92 & \\
\hline
 7 & 120 & -2981.98 & -2981.95 \\
 8 & 165 & -3139.40 & -3139.41 \\
\hline \hline
\end{tabular}
\caption{Total energies for tetrahedral fcc Cu clusters} \label{tbl:fcc_Cu}
\end{center}}
\end{table}

The energies predicted using the algebraic expression are in
excellent agreement with the `exact' calculations from first
principles as evident from the results for $\ell=7$ and $\ell=8$
tetrahedrons in Table~\ref{tbl:fcc_Cu} and Fig.~\ref{fig:cu_etot}.
The above result clearly provides further support for our method of estimation.

\subsection{Tetrahedral, zincblende based $\rm CdSe$ clusters\label{sect:znbl}}

The second system in our discussion is a non-stoichiometric,
zincblende CdSe cluster bounded by four equivalent (111) facets
terminated by Cd atoms.
As described in the Methodology section and Fig.~\ref{fig:cdse_zb},
the characteristic size is $\ell L_0$ with $L_{0} = 4.342 \AA $.
The bulk values, $E_{tot}^{bulk}(Cd)$ and $E_{tot}^{bulk}(Se)$,
are obtained from total energy calculations
of pure Cd (in hcp structure) and pure Se (in trigonal structure) separately,
while $E_{tot}^{bulk}(CdSe)$ is the total energy of a
CdSe pair (in zincblende structure). Through explicit calculations, we obtain
$E_{tot}^{bulk}(Cd) = -1467.17~eV; E_{tot}^{bulk}(Se) =
-256.97~eV; E_{tot}^{bulk}(CdSe) = -1724.84~eV.$ Therefore, the
formation energy is calculated to be $\Delta H_{f} = -0.7~eV$ (from Eq.~(\ref{eq:formation})).

\begin{figure}[htp]
 \begin{center}
  \includegraphics[width=21pc]{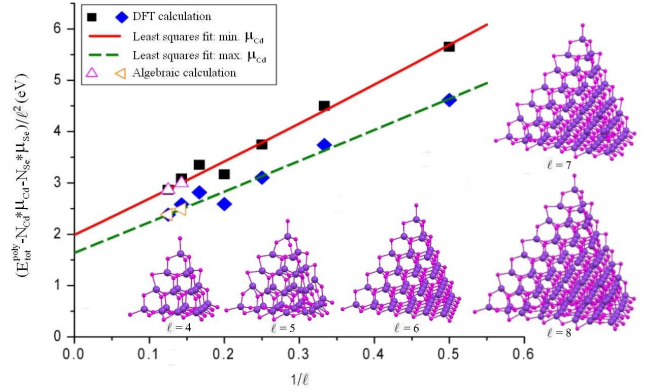}
   \caption{\small The termination energy contribution (see text)
   divided by $\ell^{2}$ for CdSe clusters as a
   function of the inverse of the characteristic size.
  The squares correspond to DFT energies while the lines are least squares fits.
  The solid line corresponds to the minimal $\mu_{Cd}$,
  while the dashed line corresponds to the maximal $\mu_{Cd}$.
  The open triangles represent energies obtained from our algebraic calculation after fitting.}
 \label{fig:cdse_etot}
 \end{center}
\end{figure}

In Table~\ref{tbl:zb_cdse}, DFT based total energies for polyhedrons
from $\ell=2$ to $\ell=8$ are shown. Using DFT based
small clusters energies (for $\ell=2, 3, 4, 5, 6$), we
have obtained the necessary fitting parameters involving surface energy, edge energy,
and corner energy,  which vary within
the intervals $50-61~meV/\AA^{2}$, $227-268~meV/\AA$, and $39-214~meV$ respectively.
We also see that certain clusters undergo noticeable reconstructions, while others do not.
Surface energy varies within the interval $50-61~meV/\AA^2$,
showing an average of surface energies within the reconstructed (for example $\ell=4,5$)
and non-reconstructed (for example $\ell=6$) structures, which is reasonably lower
than previous DFT calculations( $\sigma \sim 75-95~meV/\AA^{2}$ ~\cite{Manna}).
Surface reconstructions have been studied before using various methods, such as
the tight binding method~\cite{Pokrant}.
Here we have observed size dependent, surface reconstruction of the CdSe system
by using DFT combined with the local density approximation.

In Fig.~\ref{fig:cdse_etot}, for different $Cd$ chemical potentials, we plot the
scaled termination energy, $(E_{tot}^{poly} - N_{Cd}\mu_{Cd} -N_{Se}\mu_{Se})/\ell^{2}$
by  subtracting the  chemical potential contributions from the total energy of
the cluster as a function of $1/\ell$. The algebraic values of
$E_{tot}^{poly}$ obtained from Eq.~(\ref{eq:etot_tetra})
for $\ell =7, 8$, are indicated on the fitted curves, along with other values resulting from
direct DFT calculations.
In general the termination energy depends on the
choice of $\mu_{Cd}$, as evident from Fig.~\ref{fig:cdse_etot},
which is required to compare these surfaces with other surfaces.
However, the predicted values for the total
energy ($E_{tot}^{ploy}$) of the $\ell=7, 8$ clusters
relative to other polyhedra with the same shape can be found using Eq.~(\ref{eq:cdse}),
which yields the same value for {\sl different} values of $\mu_{Cd}$
(see Table~\ref{tbl:zb_cdse}; i.e., the algebraic value
$E^{poly}_{tot}/\ell^{2}=-4032.30 eV$ for the $\ell=7$ polyhedron
for all the values of $\mu_{Cd}$.)
This value is comparable to the energy obtained from  the direct DFT calculations,
which is an advantage since it shows that we can obtain certain
energies independent of $\mu_{Cd}$, as discussed earlier.

\begin{table}[h]
{ \begin{center}
\begin{tabular}{||c|c|c|c|c||}
        \hline
        \hline
        \multirow{2}{*}{$\ell~(L_0)$} & \multirow{2}{*}{$N_{Cu}$} & \multirow{2}{*}{$N_{Se}$} & \multicolumn{2}{c||}{$E^{poly}_{tot}/\ell^{2}~(eV)$} \\
        \cline{4-5}
                                                                       & & &DFT&algebraic \\
        \hline
       2& 10 & 4  & -3920.98 &  \\
       3& 20 & 10 & -3542.94 & \\
       4& 35 & 20 & -3528.42 &  \\
       5& 56 & 35 & -3644.61 &  \\
       6& 84 & 56 & -3821.40 &  \\ \hline
       7& 120 & 84 & -4032.21  & -4032.30  \\
       8& 165 & 120 & -4263.30 & -4263.30 \\ \hline \hline
\end{tabular}
\caption{CdSe (zincblende) total energies for various cluster sizes. } \label{tbl:zb_cdse}
\end{center}}
\end{table}

\section{Discussion}

For polyhedral nanoclusters of fcc Cu, calculations  from
10 to 165 atoms show that the energies are well described by this form
even for small clusters. Thus the energies for all sizes can be determined
efficiently based on calculations for small clusters and
we propose that this is a useful approach for metals.

When testing this approach for CdSe polyhedrons that are Cd terminated
and non-stoichiometric, one encounters energy contributions
that depend on the individual chemical potentials, as seen from Eq.~(\ref{eq:etot})
which defines $E_{tot}^{poly}$.
These chemical potentials
are subject to upper and lower bounds\cite{Martin}
(see inequalities~(\ref{eq:bounds1}),~(\ref{eq:bounds2})).
However, it turns out that the total energies of even
the non-stoichiometric CdSe polyhedra considered here
can be obtained without knowing the individual chemical potentials.
This is sufficient to extrapolate to large size for this family
of clusters independent of chemical potentials and using only directly
calculated total energies.
The reason for the above is that the total energy of such clusters
can be calculated from well defined bulk energies  by adding surface, edge and corner
termination terms; these termination terms appear as mere parameters that
scale with the dimensions of the cluster and can be estimated from DFT calculations
of small clusters with similar topologies.
In addition, the surface, edge and corner
termination terms can be used for other clusters by  including the
chemical potentials in the way given in Eq.~(\ref{eq:CancelVolTerm}).

For CdSe, we find an overall trend similar to that for Cu;
however, there are deviations from a smooth curve for
the energies as a function of size.
The deviations are likely to be associated with reconstructions
that are seen for certain nanocluster sizes. These reconstructions probably
originate from the changes in state occupations  near the
Fermi energy, with the largest changes apparently occurring for the edges.
Also note that as the cluster size $L\to \infty$, the corner and edge contributions
become small compared to the surface energies and
thus providing a way of estimating the latter for large clusters.

The total energies are described well by a least squares fit, carried out using
the results for 5 clusters with 14 to 140 atoms, which accurately
determines the energies for clusters with 204 and 285 atoms.
However, the division into three separate contributions as
surface, edge and corner energies, is not as well determined
due to the (possible reconstruction-induced) variations between the different clusters.
The estimated surface energy represents an average of different
reconstructed and unreconstructed clusters, which is reasonably
lower than the value reported in previous DFT calculations, as pointed out in
the results section.

\section{Conclusion}

In summary, we have presented an efficient method for calculating
total energy  of large
nanoclusters using a parametrized, algebraic form with parameters fitted from small,
first principles based, nanocluster calculations. The method appears to work
quite well for pure metals. Even for non-stoichiometric, semiconducting
clusters with polar surfaces, this approach provides a way of estimating
total energies of large clusters,
important information such as relaxation energies
that are specific to the chosen clusters, as well as
surface, edge and corner energies that can be used for other clusters.


\begin{thebibliography}{99}

\bibitem{Murray}
C. B. Murray, C. R. Kagan, and M. G. Bawendi,
Annu. Rev. Mater. Sci. {\bf 30}, 545 (2000)

\bibitem{Alivisatos}
A. P. Alivisatos,
J. Phys. Chem. {\bf 100}, 13226 (1996)

\bibitem{Skolnick & Mowbray}
M. S. Skolnick, and D. J. Mowbray,
Annu. Rev. Mater. Res. {\bf 34}, 181 (2004)

\bibitem{Law}
M. Law, J. Goldberger, and P. D. Yang,
Annu. Rev. Mater. Res. {\bf 34}, 83 (2004)

\bibitem{Tiller}
W. A. Tiller,
The Science of Crystallization: Microscopic Interfacial Phenomena (Cambridge University Press, Cambridge, 1991)

\bibitem{Peng}
X. G. Peng, L. Manna, W. D. Yang, J. Wickham, E. Scher, A. Kadavanich, and A. P. Alivisatos,
Nature {\bf 404}, 59 (2000).

\bibitem{Nirmal}
M. Nirmal, D. J. Norris, M. Kuno, M. G. Bawendi, Al. L. Efros, and M. Rosen,
Phys. Rev. Lett. {\bf 75}, 3728 (1995).

\bibitem{Cleveland & Landman}
C. L. Cleveland, and U. Landman,
J. Chem. Phys. {\bf 94}, 7376 (1991).

\bibitem{Yacaman}
M. J. Yacaman, J. A. Ascencio, H. B. Liu, and J. G. Torresdey,
J. Vac. Sci. Technol. B. {\bf 19}, 1091 (2001).

\bibitem{Zhang}
 S. B. Zhang and S. H. Wei,
 Phys. Rev. Lett. {\bf 92}, 086102 (2004).

\bibitem{Manna}
 L. Manna, L. W. Wang, R.Cingolani, and A. P. Alivisatos,
 J. Phys. Chem. B {\bf 109}, 6183 (2005).

\bibitem{Bernholc}
 K. Rapcewicz, B. Chen, B. Yakobson, and J. Bernholc,
 Phys. Rev. B {\bf 57}, 7281 (1998).

\bibitem{Martin}
R. M. Martin,
Electronic Structure: Basic Theory and Practical Methods (Cambridge University Press, New York, 2004)

\bibitem{Siesta}
J. M. Soler et al.,
J. Phys.: Condens. Matter, {\bf 14}, 2745 (2002).

\bibitem{Troullier & Martins}
N. Troullier and J. L. Martins,
Phys. Rev. B 43, 1993 (1991).

\bibitem{Scheffler}
 H. M. Polatoglou, M. Methfessel, and M. Scheffler,
 Phys. Rev. B {\bf 48}, 1877 (1993).

\bibitem{Boer}
 F. R. de Boer, R. Boom, W. C. M. Mattens, A. R. Miedema, and A. K. Niessen,
 {\em Cohesion in Metals}(North-Holland, Amsterdam)(1988).

\bibitem{Pokrant}
S. Pokrant, and K. B. Whaley
Eur. Phys. J. D. {\bf 6} 255 (1999).

\end{thebibliography}
\end{document}